\documentclass{aastex61}
\usepackage{natbib}

\begin{document}
\title{Parametric Decay Instability and Dissipation of Low-frequency Alfv\'en Waves in Low-beta Turbulent Plasmas}
\shorttitle{PDI in Low-beta Turbulent Plasmas}
\shortauthors{Fu et al.}
  
\author{Xiangrong Fu}
\email{sfu@newmexicoconsortirum.org}
\affiliation{New Mexico Consortium, Los Alamos, NM 87544}
\author{Hui Li}
\affiliation{Los Alamos National Laboratory, Los Alamos, NM 87545}
\author{Fan Guo}
\affiliation{Los Alamos National Laboratory, Los Alamos, NM 87545}
\author{Xiaocan Li}
\affiliation{Los Alamos National Laboratory, Los Alamos, NM 87545}
\author{Vadim Roytershteyn}
\affiliation{Space Science Institute, Boulder, CO 80301}
\keywords{Alfv\'en wave; parametric decay; slow mode; hybrid simulation}

\begin{abstract}
  Evolution of the parametric decay instability (PDI) of a
circularly polarized Alfv\'en wave in a turbulent low-beta plasma background is
investigated using 3D hybrid simulations.  It is shown
that the turbulence reduces the growth rate of PDI as compared to the linear
theory predictions, but PDI can still exist. Interestingly, the damping rate of
ion acoustic mode (as the product of PDI) is also reduced as compared to the
linear Vlasov predictions. Nonetheless, significant heating of
ions in the direction parallel to the background magnetic field is observed due
to resonant Landau damping of the ion acoustic waves.  In low-beta turbulent plasmas, PDI can provide an important
channel for energy dissipation of low-frequency Alfv\'en waves at a scale much larger than the ion
kinetic scales, different from the traditional turbulence dissipation models.
\end{abstract} 

\section{Introduction}

An Alfv\'en wave is a fundamental magnetohydrodynamic (MHD) mode that prevails
in laboratory, space and astrophysical plasmas.  In solar corona and solar wind, Alfv\'en waves are ubiquitous as
a major carrier of fluctuating magnetic energy. They are often observed to have
large amplitudes so that nonlinear wave-wave and wave-particle interactions are
expected to be important \citep{tu_ssr_1995}.

It is rare to observe narrow band Alfv\'en waves in the solar wind. Instead,
measured magnetic fluctuations are mostly turbulent, in the sense that the
energy spreads across a wide range of frequency, typically over 3-4 orders of
magnitude. This is believed to be a result of nonlinear wave-wave interaction
among counter-propagating Alfv\'en packets \citep{gs95}, where long-wavelength perturbations are
cascading into short-wavelength ones and form a power-law energy spectrum from
the injection scale down to ion kinetic scales ($\rho_i$ or $d_i$)
\citep[e.g.,][]{howes_ptrsa_2015}. 

On the other hand, in compressible plasmas large amplitude Alfv\'en waves (and magnetosonic waves) are
subject to a parametric decay instability (PDI) -- another category of nonlinear
wave-wave interactions -- where a forward propagating Alfv\'en wave (pump wave)
decays into a backward propagating Alfv\'en wave and a forward propagating ion
acoustic wave or slow wave. (Note that in the paper, we use the terms ``slow
mode'' and
``ion acoustic mode'' interchangeably in low-beta plasmas,
see \citet{gary_book_1993} and \citet{versc_apj_2017} for relations between
these two modes.) PDI provides not only
a robust mechanism for generating backward propagating waves, which are a key
ingredient in the turbulence cascading process mentioned above, but also a
mechanism to dissipate wave energy into plasma through Landau damping of ion acoustic
waves. This dissipation of energy can be efficient because it occurs at a fluid
scale (comparable to wavelength of Alfv\'en waves, $kd_i \ll 1$), in contrast to dissipation
at ion kinetic scales where the energy budget is much smaller. 

\begin{figure}
   \plotone{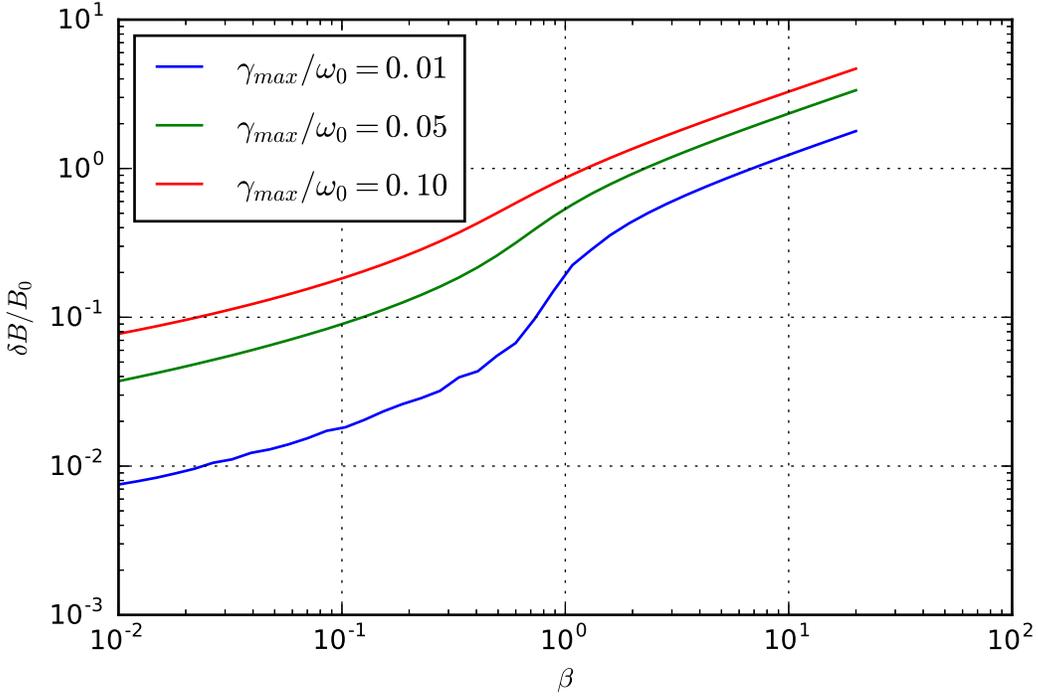}
   \caption{The dependence of pump wave amplitude
    on the plasma beta for various maximum growth rates of PDI
    $\gamma_{\rm max}/\omega_0=0.01, 0.05, 0.10$.\label{fig:linear}}
 \end{figure}

There have been comprehensive theoretical and simulation studies on parametric
decay instabilities of Alfv\'en waves in quiescent plasmas in the literature.
PDI of linearly polarized Alfv\'en wave in the limit of small pump wave
amplitude ($\delta
B/B_0$) and low
plasma beta ($\beta$) was studied by \citet{sagdeev_book_1969}, and its growth
rate is 
\begin{equation}
  \frac{\gamma}{\omega_0}\approx \frac{1}{2}\frac{\delta
  B}{B_0}\frac{1}{\beta^{1/4}},
  \label{eq:pdi_growth}
\end{equation}
where $\omega_0$ is the frequency of the pump wave.
For a circularly
polarized Alfv\'en (CPA) wave, the dispersion relation in the MHD limit can be
obtained for
finite $\delta B/B_0$ and $\beta$, as given independently by \citet{derby_apj_1978} and
\citet{golds_apj_1978}.
The dependence of the pump wave amplitude on plasma beta for various maximum
growth rates of PDI $\gamma_{\rm max}/\omega_0=0.01, 0.05, 0.10$ (by solving
Eq.~17 of \citet{derby_apj_1978}) are shown in
Figure \ref{fig:linear}. It is clear that the
threshold amplitude of PDI is low in low-beta plasmas, e.g., an Alfv\'en wave with amplitude as
small as $10^{-2}$ will be unstable when $\beta<10^{-2}$.

The nonlinear features of PDI have been investigated by numerical simulations,
including MHD and hybrid simulations, for monochromatic\citep{gao_pop_2013b} and
nonmonochromatic Alfv\'en waves\citep{malara_pop_2000}, at parallel and oblique
propagation\citep{zanna_grl_2001,matte_grl_2010}, in multiple
dimensions\citep{matte_grl_2010, gao_pop_2013}, with multiple ion
species\citep{gao_pop_2013b}, and in expanding solar wind
\citep{tener_jgr_2013,zanna_jpp_2015}. These studies focused on PDI in quiescent
plasmas. PDI in turbulent plasmas has not been explored until recently.  Using
3D MHD simulations, \citet{shi_apj_2017} found that PDI can survive turbulence
over a broad range of parameters, only with growth rates reduced by about 50\%.
But kinetic effects were missing in their study due to the limitation of MHD
model. Can PDI survive in a more realistic turbulent environment?  In this
study, we address this question using 3D hybrid simulations with full ion kinetics
while still retain large-scale turbulence in the inertial range. In a turbulent
warm plasma, the growth and damping of various plasma waves predicted by the
linear Vlasov theory may be altered significantly, challenging the effectiveness
of parametric decay on ion heating of ions, which is also addressed in the
current paper.

A main evidence of PDI in space plasmas is the presence of slow modes in in-situ measurements
\citep[][and references therein]{tu_ssr_1995}. \citet{spang_pop_1997} reported signature in the field and
density spectra consistent with the PDI theory in the upstream of the Earth's
bow shock. Slow modes have been identified in various analyses of solar wind data
\citep[e.g.][]{kello_angeo_2005, howes_apj_2012, yao_apj_2013}. Recently, \citet{shi_apj_2015} analyzed data from WIND spacecraft in
Solar Cyle 23, and found existence of slow waves in 3.4\% of the selected time
period, preferentially in moderate-speed solar wind. However, in general nonlinear wave-wave interactions are difficult to
pinned down \citep{narita_npg_2007}.  In laboratory
experiments, a parametric instability of finite $\omega/\Omega_i$ kinetic
Alfv\'en wave (modulational instability) was directly observed for the first time on the Large Plasma Device
recently \citep{dorfm_prl_2016}; but verification of the classic PDI remains challenging. 

\section{Simulation Model}

We use hybrid simulation (kinetic ions and mass-less fluid electrons) to study
the parametric decay of large amplitude Alfv\'en waves in low-beta turbulent
electron-proton plasmas. The hybrid model is appropriate because we focus on
energy dissipation at long wavelength ($k d_i\ll 1$). A 3D hybrid code H3D
\citep{karim_asp_2006, podes_jgr_2017} is used in this study with the following
typical parameters:
$L_x=L_y=L_z/4=120 d_i$, $N_x=N_y=N_z=480$, $\Delta t\Omega_i=0.025$, 
$\beta_i=0.01$. The simulation box is elongated to facilitate development of anisotropic
($k_\perp\gg k_\parallel$) MHD turbulence \citep{gs95}. Typically 64 marker
particles in each cell are used to represent protons. To ensure that our results are not
sensitive to the numerical noise, selected simulations with higher number of
particles per cell (NPPC) are also
carried out. The plasma is immersed in a uniform background magnetic field ${\bf
B}_0=B_0{\bf z}$. Periodic boundary conditions for fields and particles are
applied. Electrons have the same temperature as ions initially, and follow an adiabatic
equation of state $T_e/n_e^{\Gamma-1}=\rm const$, where $\Gamma=5/3$. A small
uniform resistivity $\eta_e=10^{-8}4\pi/\omega_{pi}$ is assumed. Total
energy is conserved within a few percents in all simulations presented.
Key parameters for our 3D hybrid simulations are summarized in Table
\ref{tab:para}.

 \begin{table}
   \centering
   \begin{tabular}{c|c|c|c|c|c|c}
     Run &number of cells  &$\beta_i$& $a_0$ &$a_1$&
     $t_1\Omega_i$&NPPC\\\hline\hline
     0 & $480\times480\times 480$& 0.01&0.0&0.1&0 &64\\\hline
     1 & $480\times480\times 480$& 0.01&0.1&0.1&500&64\\\hline
     2 & $480\times480\times 480$& 0.01&0.1&0.0&N/A&64\\\hline
     3 & $480\times480\times 480$& 0.01&0.1&0.1&500&216\\\hline
     4 & $120\times120\times 480$& 0.01&0.1&0.1&500&64\\\hline
     5 & $120\times120\times 480$& 0.3&0.3&0.3&500&64\\\hline
     6 & $120\times120\times 480$& 0.3&0.3&0.0&N/A&64\\
   \end{tabular}
   \caption{Key parameters for 3D hybrid simulations.  The domain size is $120\times120\times 480 d_i^3$ for all runs. $a_0$ is the
   amplitude of large scale Alfv\'en waves producing background turbulence. $a_1$ is the
 amplitude of injected Alfv\'en wave and $t_1$ is the injection time. NPPC is
 number of particles per cell.}
   \label{tab:para}
 \end{table}

At t=0, three pairs of
counter-propagating long-wavelength Alfv\'en waves are loaded throughout the simulation
domain so that the 
fluctuating magnetic and velocity fields are \citep{makwa_pop_2015}:
\begin{eqnarray}
  \delta {\bf B}/B_0&=&\sum_{j,k}a_0\cos(jk_0 y+kk_0
  z+\phi_{j,k})\hat{\bf x}\nonumber\\
  &&+\sum_{l,n}a_0\cos(lk_0 x +n k_0 z+\phi_{l,n})\hat{\bf y}
\end{eqnarray}
\begin{eqnarray}
  \delta {\bf v}/v_A&=&\sum_{j,k}a_0{\,\rm sgn}(k)\cos(jk_0 y+kk_0
  z+\phi_{j,k})\hat{\bf x}\nonumber\\
  &&+\sum_{l,n}a_0{\,\rm sgn}(n)\cos(lk_0 x+n k_0 z+\phi_{l,n})\hat{\bf y}
\end{eqnarray}
where  $(j,k)= (4,1); ( 8,1); ( 12, -2)$, $(l,n)= (4,-1);(-8,
-1);(-12, 2)$ and $k_0=2\pi/L_z=0.013d_i^{-1}$.  Nonlinear interactions of these waves allow
energy cascading to form a power-law turbulent spectrum \citep{gs95,howes_ptrsa_2015}. After the turbulence has been
established, at $t=t_1$ ($t_1>\tau_A$, where $\tau_A \equiv
L_z/v_A$ is the Alfv\'en transit time) a circularly polarized Alfv\'en wave propagating
along the background magnetic field,  with $\delta {\bf B}/B_0=a_1\cos(k_1
z)\hat{\bf x}+a_1\sin(k_1 z)\hat{\bf y}$ and $\delta {\bf v}/v_A=-\delta {\bf
B}/B_0$, is injected (superposed on the existing fields) to excite
PDI.  For a circularly polarized wave, the magnetic pressure $|{\bf B}|^2/8\pi$ is
spatially uniform and no ponderomotive force is exerted.

\section{Results}
\begin{figure}
   \centering
   (a)\hspace{0.45\textwidth}(b)\\
   \includegraphics[width=0.45\textwidth]{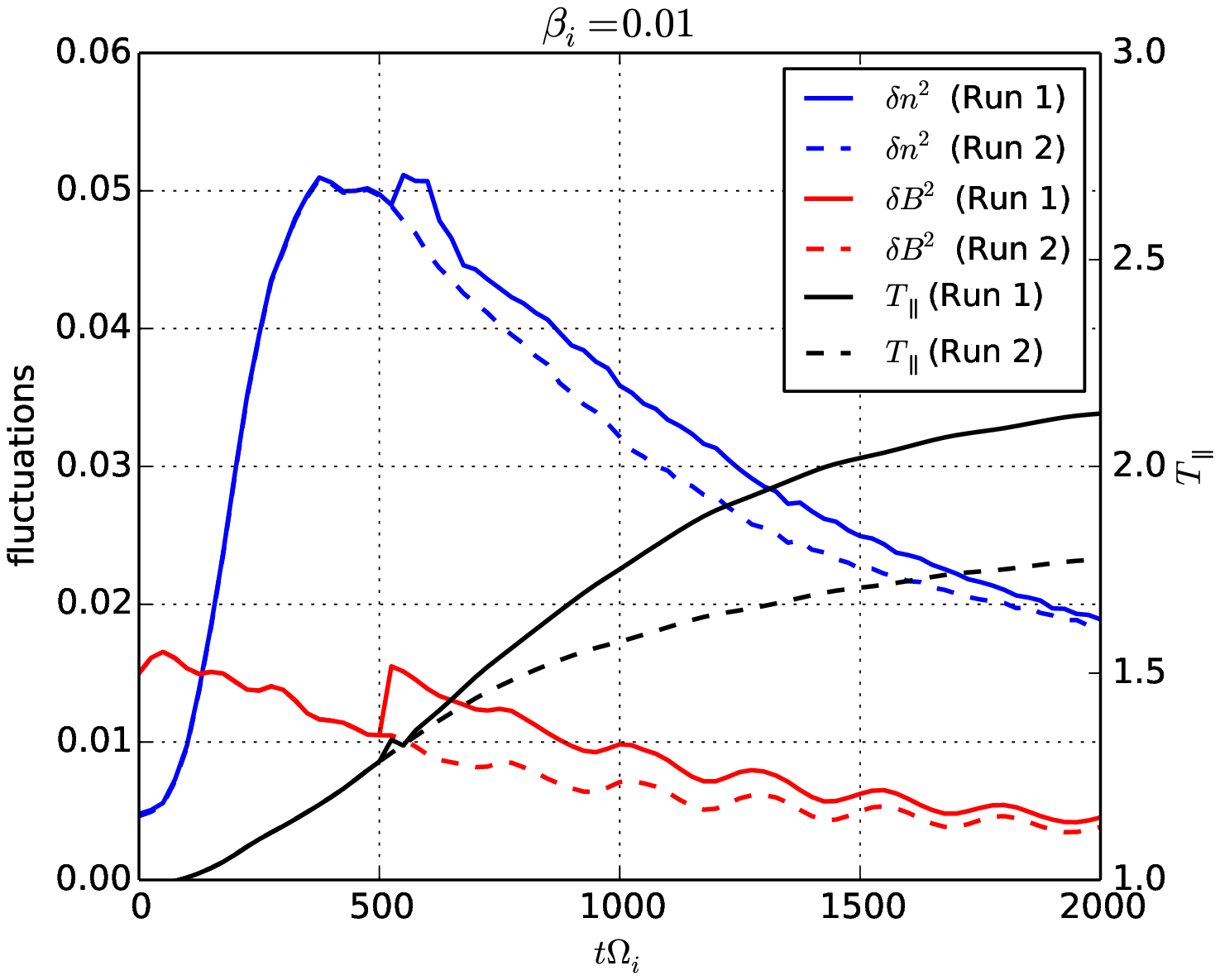}
   \includegraphics[width=0.45\textwidth]{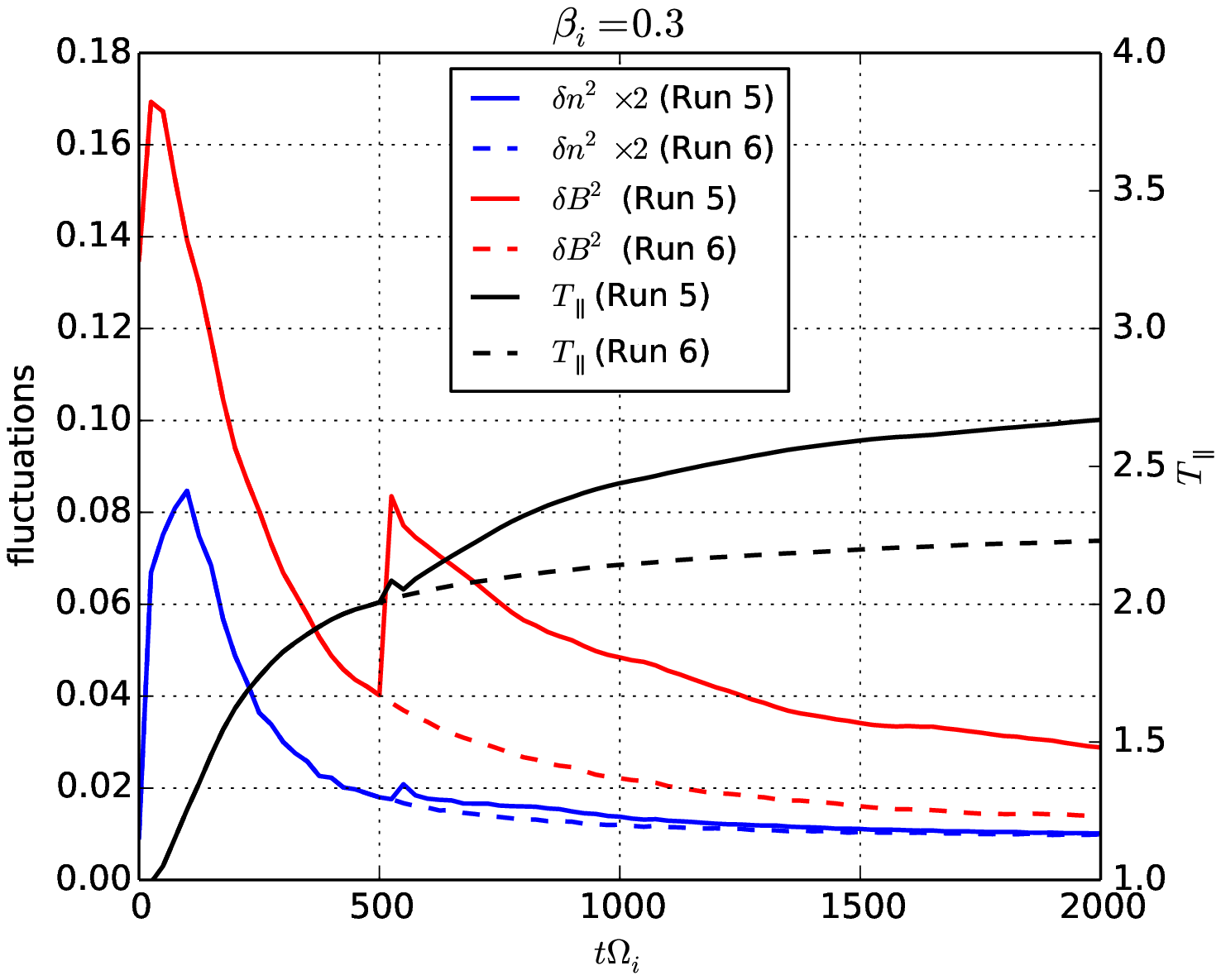}
   \caption{Time evolution of magnetic field fluctuation $(\delta B/B_0)^2$,
     plasma density fluctuation $(\delta n/n_0)^2$ and parallel ion
     temperature $T_\parallel/T_{\parallel 0}$ in (a) Run 1 and Run 2 with
     $\beta_i=0.01$
     and (b) Run 5 and Run 6 with $\beta_i=0.3$. In Run 1 and Run 5 a circularly polarized Alfv\'en wave is injected at
     $t\Omega_i=500$, and no wave is injected in Run 2 or Run 6 after the simulation starts.}
 \label{fig:history}
 \end{figure}

An overview of simulation Run 1 is shown in Figure \ref{fig:history}a (solid
lines).
We start with six Alfv\'en waves each of which has an amplitude
$a_0=0.1$. These
waves interact with each other causing energy cascade, establishing a
background turbulence. 
Analyses of magnetic field fluctuation (not shown) reveal a power-law spectrum in the
direction perpendicular to the background magnetic field with
$\delta B^2\propto k_\perp^{-2}$  extending from $k_\perp d_i=0.2$ to
$k_\perp d_i=2$ within one Alfv\'en time ($\tau_A=480 \Omega_i^{-1}$).
The wave power in the parallel direction has a much steeper
spectrum, roughly proportional to $k_\parallel^{-2.7}$ in the
range $0.04<k_\parallel d_i <0.4$, probably due to preferential cascading of MHD
turbulence in the perpendicular direction
\citep{gs95,howes_ptrsa_2015,ought_ptrsa_2015}.
During the process, the magnetic and
electric field energies are decreasing and ions are being heated in both
parallel and perpendicular directions, similar to
previous MHD and full-PIC simulation results
\citep{makwa_pop_2015,makwa_arx_2016,zhdan_prl_2017}. The density fluctuation
also grows and it saturates at $t\Omega_i\approx400$.  
At $t=500\Omega_i^{-1}$ , we inject an
circularly polarized Alfv\'en wave with an amplitude $a_1=0.1$ and wavelength
$k_1d_i=0.13$ into the system. The injection causes an abrupt increase of
field energies around $t\Omega_i=500$. According to the theory of PDI, an slow
mode with density perturbation will be excited and the presence of such mode is
evident as the jump in density fluctuation around $t\Omega_i=500$.  For
comparison, results of Run 2, which is identical to Run 1 except that no wave is
injected after the simulation starts, are shown in dashed curves in Figure
\ref{fig:history}a. Although the final turbulence level is close to that of Run
1, parallel ion temperature at the end of Run 2 is significantly smaller than
that of Run 1. 

\begin{figure}
   \centering
   \includegraphics[width=0.45\textwidth]{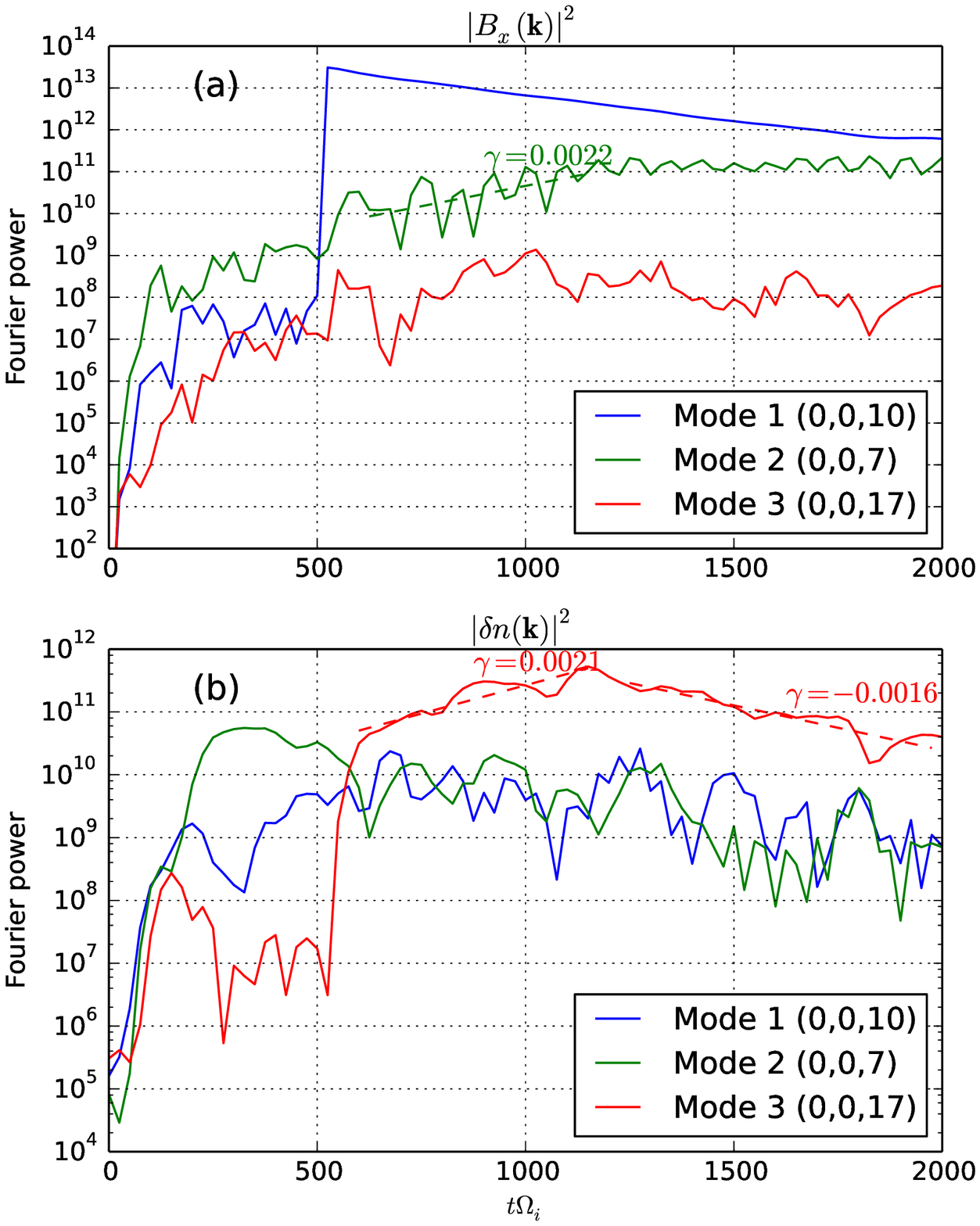}
   \includegraphics[width=0.45\textwidth]{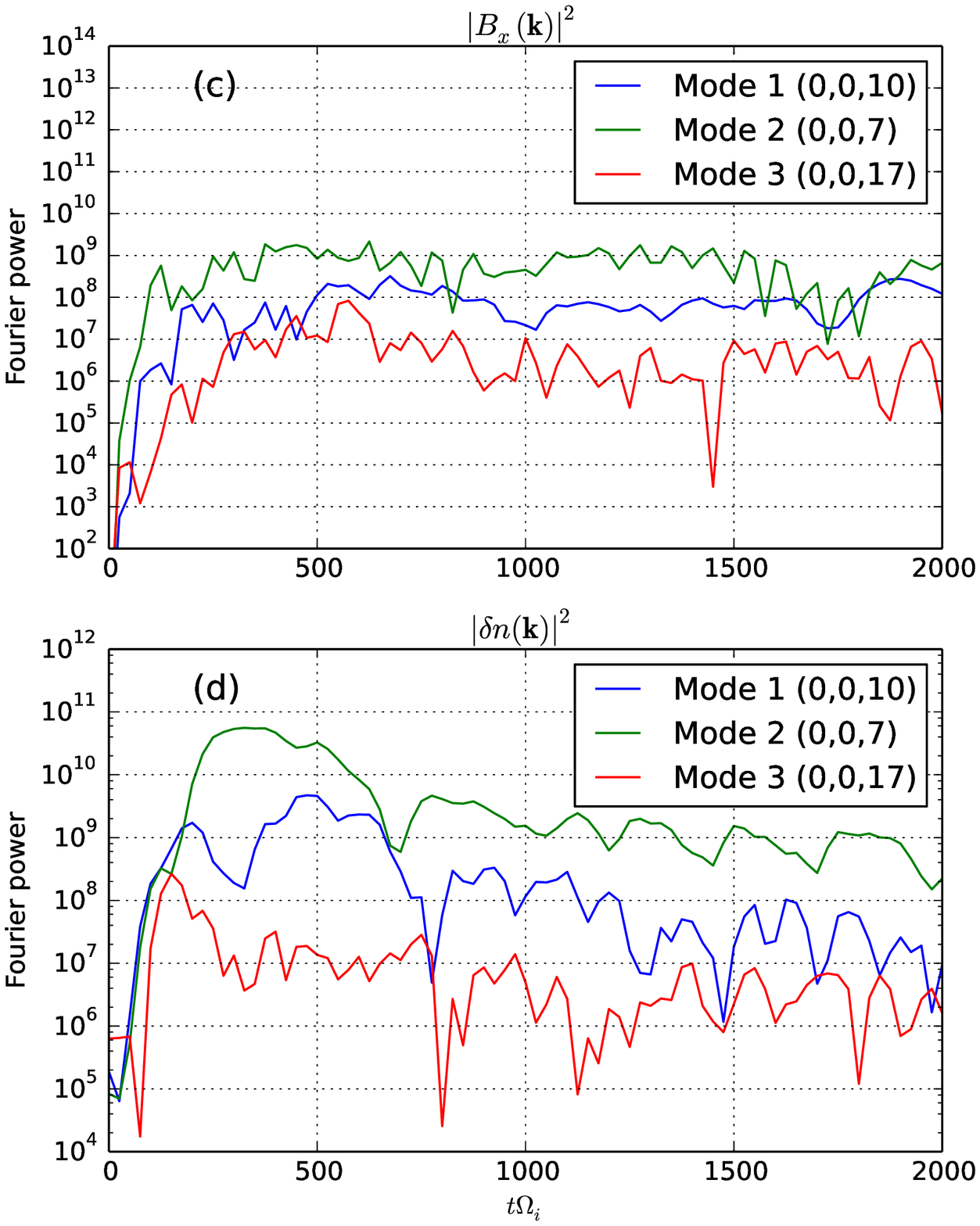}
   \caption{Time evolution of wave power of (a) magnetic field $B_x$ and
     (b) density $\delta n$ in Run 1 for the three modes of interest $(k_x,k_y,k_z)/k_0=(0,0,7); (0,0,10); (0,0,17)$.
   As a comparison, time evolution of wave power of $B_x$ and $\delta n$ in Run
   2 (no injection) for the same three modes are shown in
   panel (c) and (d), respectively.
 }
 \label{fig:mode}
 \end{figure}

In 3D simulations, PDI is free to develop in parallel and oblique
directions, as shown in the previous study \citep{shi_apj_2017}. Here we focus
on three modes of special interest in Run 1, i.e. modes with
$(k_x,k_y,k_z)/k_0=(0,0,10), (0,0,7)$ and $(0,0,17)$ based on the linear theory
of parallel PDI \citep{derby_apj_1978,golds_apj_1978}. Mode (0,0,10) is the forward-propagating Alfv\'en wave (hereafter
named Mode 1) injected after
the establishment of background turbulence; mode (0,0,7) and (0,0,17) are the
backward-propagating Alfv\'en wave (Mode 2) and the forward-propagating slow
mode (Mode 3) predicted
by the linear theory, respectively. Their wave frequencies and wave numbers
satisfy the three wave resonance conditions:
\begin{eqnarray}
  \omega_1& =& \omega_2+\omega_3, \\
  {\bf k}_1& =& {\bf k}_2+{\bf k}_3,
\end{eqnarray}
which are analogous to energy and momentum conservations in quantum mechanics
(if multiplied by $\hbar$).
Figures \ref{fig:mode}a and \ref{fig:mode}b show the evolution of the wave power of perpendicular magnetic
field component $B_x$ and density fluctuation $\delta n$ of these modes.
Before the injection, the magnetic field and density fluctuations of all three
modes grow and
saturate due to the energy cascades from large to small spatial scale.
After the injection, the magnetic energy of Mode 1 decreases while that of Mode
2 increases, consistent with the PDI process. The density fluctuations of both Mode 1
and Mode 2 are very small. By using dispersion relation analysis
\citep[e.g.][Figure 8]{shi_apj_2017} we confirm that Mode 2 is an Alfv\'en wave
because it is frequency ($\omega=0.088\Omega_i$) is close to $k_\parallel
v_A=0.091\Omega_i$ for $k_\parallel=7k_0$. Despite
fluctuation due to background turbulence, we fit time series of wave power of fluctuating
magnetic field or density to a exponential function $e^{\gamma t}$ to calculate growth rates. The growth rate of Mode 2 is
estimated to be $\gamma/\Omega_i=0.0022$ in the interval $600<t\Omega_i<1100$,
smaller than the PDI growth rate in a quiescent plasma, which is
$\gamma/\Omega_i=0.0088$ (measured in Run 0, a simulation without background
turbulence). The result confirms the reduction of
PDI growth rate by turbulence found in previous MHD simulations \citep{shi_apj_2017}. 
Since Mode 1 and 2 are Alfv\'en
modes, their density fluctuation levels remain nearly unchanged. Meanwhile, $\delta n$ of Mode
3 increases at a rate close to the growth rate of Mode 2
($\gamma/\Omega_i=0.0021$), but its magnetic
field has no significant growth because of the characteristics of an ion acoustic mode.  
Dispersion relation analysis also confirms that Mode 3 has frequency
$\omega=0.035\Omega_i$, close to the ion acoustic frequency $k_\parallel
c_s=0.044\Omega_i$ for $k_\parallel =17 k_0$. 
For comparison, there is no growth of density or magnetic field fluctuation
after $t\Omega_i=500$ in 
Run 2 (no injection), as shown in Figures \ref{fig:mode}c and \ref{fig:mode}d.
After $t\Omega_i=1200$, PDI saturates and the slow mode starts to decay at a rate $\gamma/\Omega_i=-0.0016$. This rate is about 1/6
of the Landau damping rate predicted by linear Vlasov theory
\citep{gary_book_1993}, which is
$\gamma_L/\Omega_i=-0.01$ based on the plasma parameters at $t\Omega_i=1200$.
In a strongly turbulent environment, the assumptions of linear theory including
unperturbed charged particle orbits and uniform plasma distribution are no longer valid.
In our simulation, strong density fluctuations shown in Figure
\ref{fig:history}a
could be the main reason why the observed damping rate is much smaller than the 
theoretical prediction.
Nevertheless, damping of the slow mode still leads to significant heating of ions in
the parallel direction,
which is also shown in Figure \ref{fig:history}a. Compared to Run 2, an additional 30\%
increase of ion temperature observed in Run 1 is due to the injection of the
Alfv\'en wave and subsequent damping of the slow mode. We obtain very similar
results in simulations with 264 particles
per cell (Run 3) and lower spatial resolution (Run 4), showing that the results
are insensitive to numerical noise and resolution.

\begin{figure}
   \centering
   \includegraphics[width=0.8\textwidth]{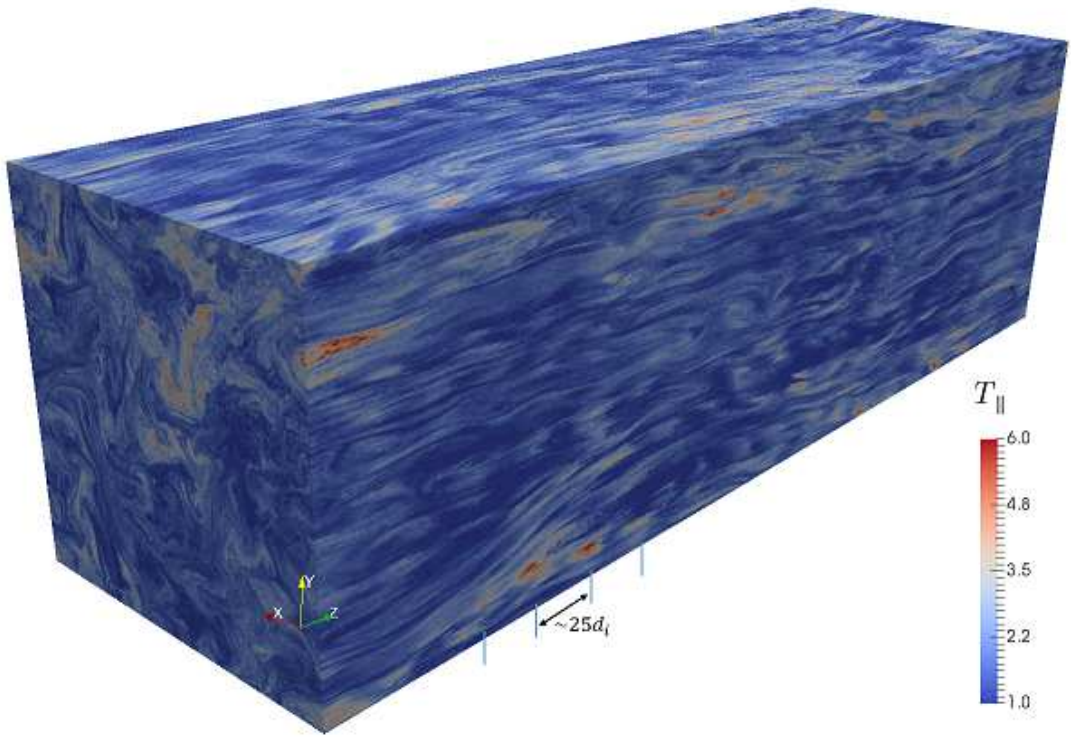}
   \caption{The surface contour plot of ion temperature $T_\parallel/T_{\parallel 0}$ at
     $t\Omega_i=1000$ in Run 1. Localized ion heating leads to temperature
     enhancement structures whose parallel size is close to the wavelength of the slow mode predicted by the PDI theory.}
 \label{fig:temp}
 \end{figure}

\begin{figure}
   \centering
   \includegraphics[width=0.45\textwidth]{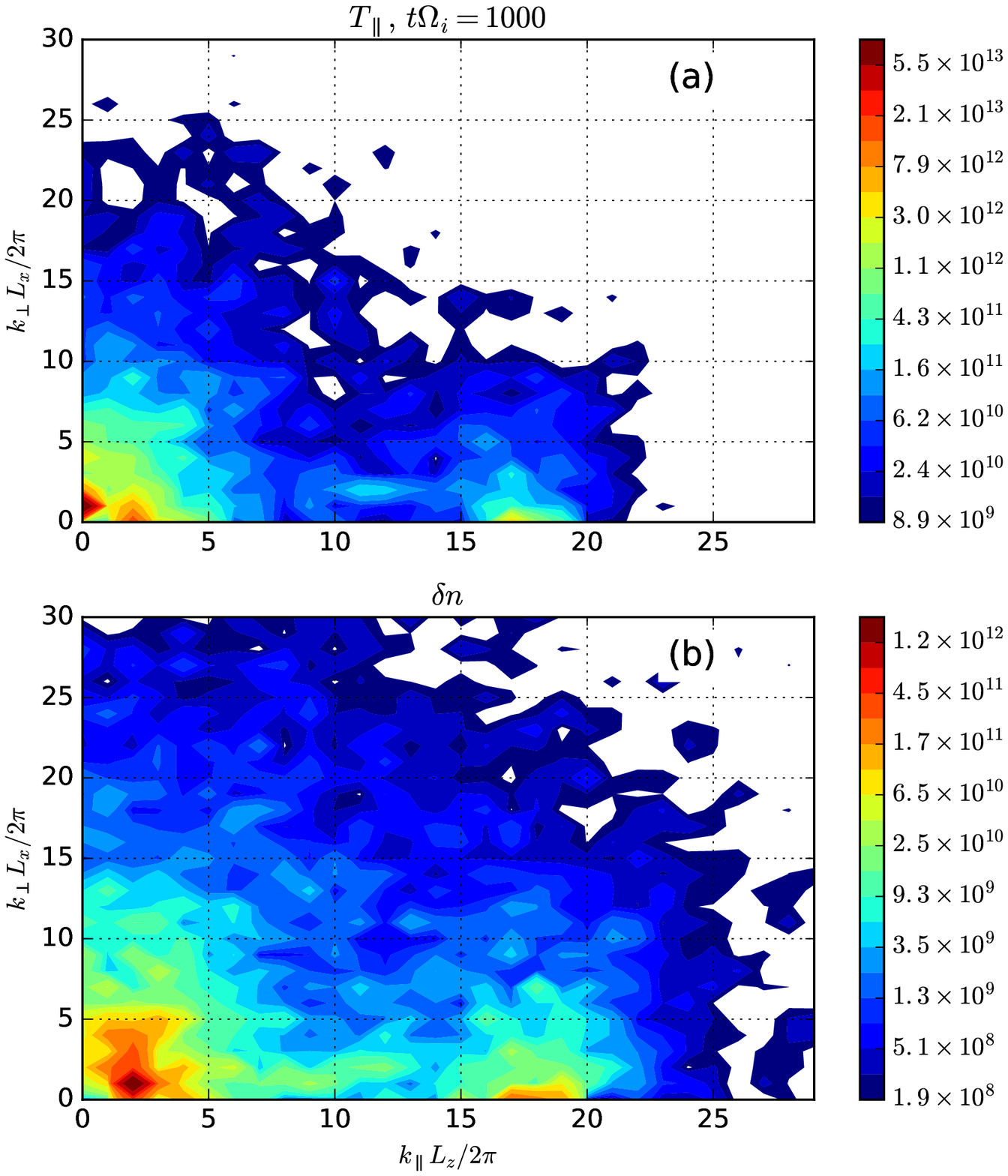}
   \includegraphics[width=0.45\textwidth]{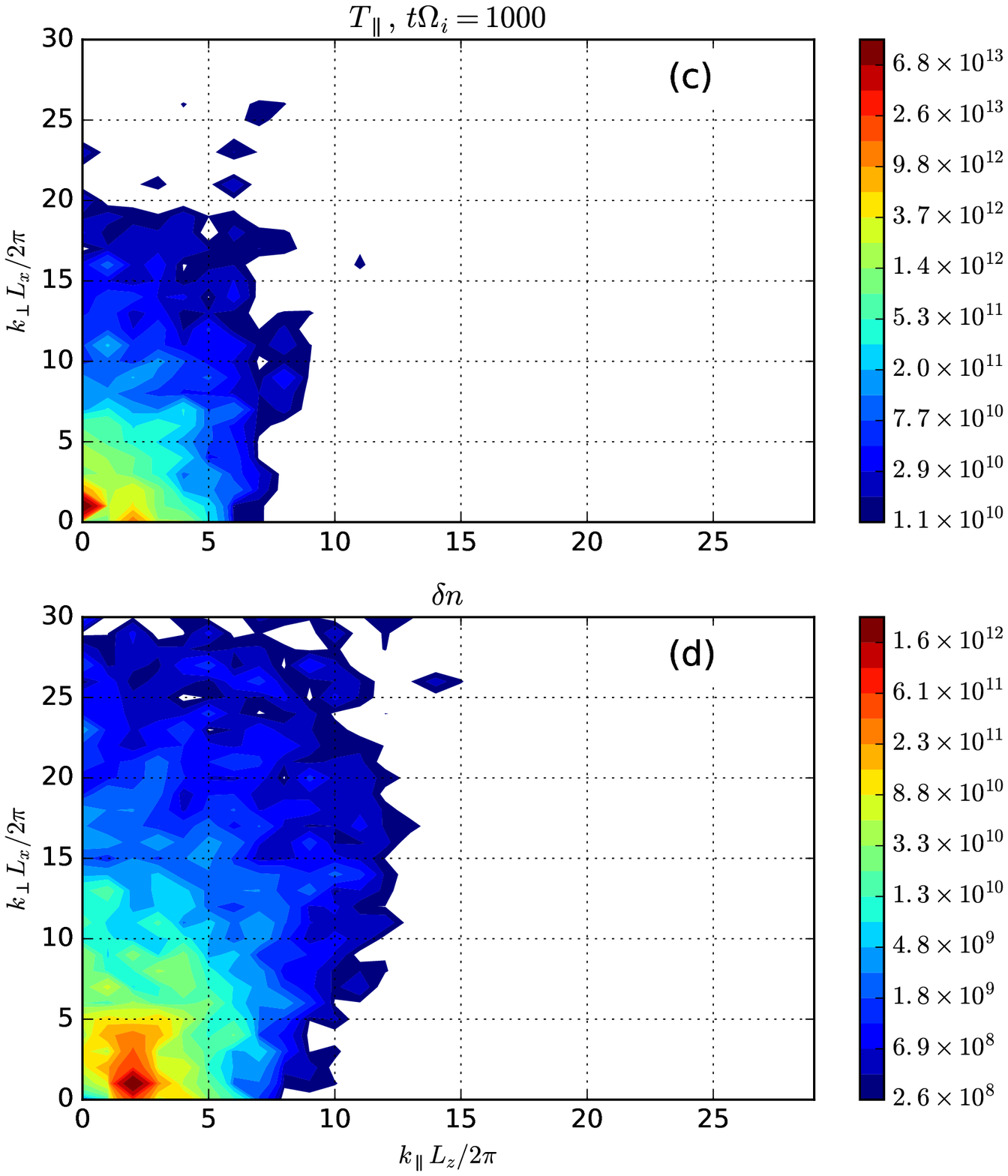}
   \caption{Power spectra of (a) parallel ion temperature $T_\parallel$ and (b) density
     fluctuation $\delta n$ as a function of $k_\parallel$ and $k_\perp$ at
     $t\Omega_i=1000$ in Run 1. Localized structures in Figure 4 corresponding to
   wave number $k_\parallel L_z/2\pi=17$, are associated with density fluctuation
 near mode (0,0,17) produced by PDI. As a comparison, power spectra of
 $T_\parallel$ and $\delta n$ at $t\Omega_i=1000$ in Run 2 (no injection) are
 shown in panel (c) and (d), respectively.}
 \label{fig:fft}
 \end{figure}

\begin{figure}
   \centering
   \includegraphics[width=0.8\textwidth]{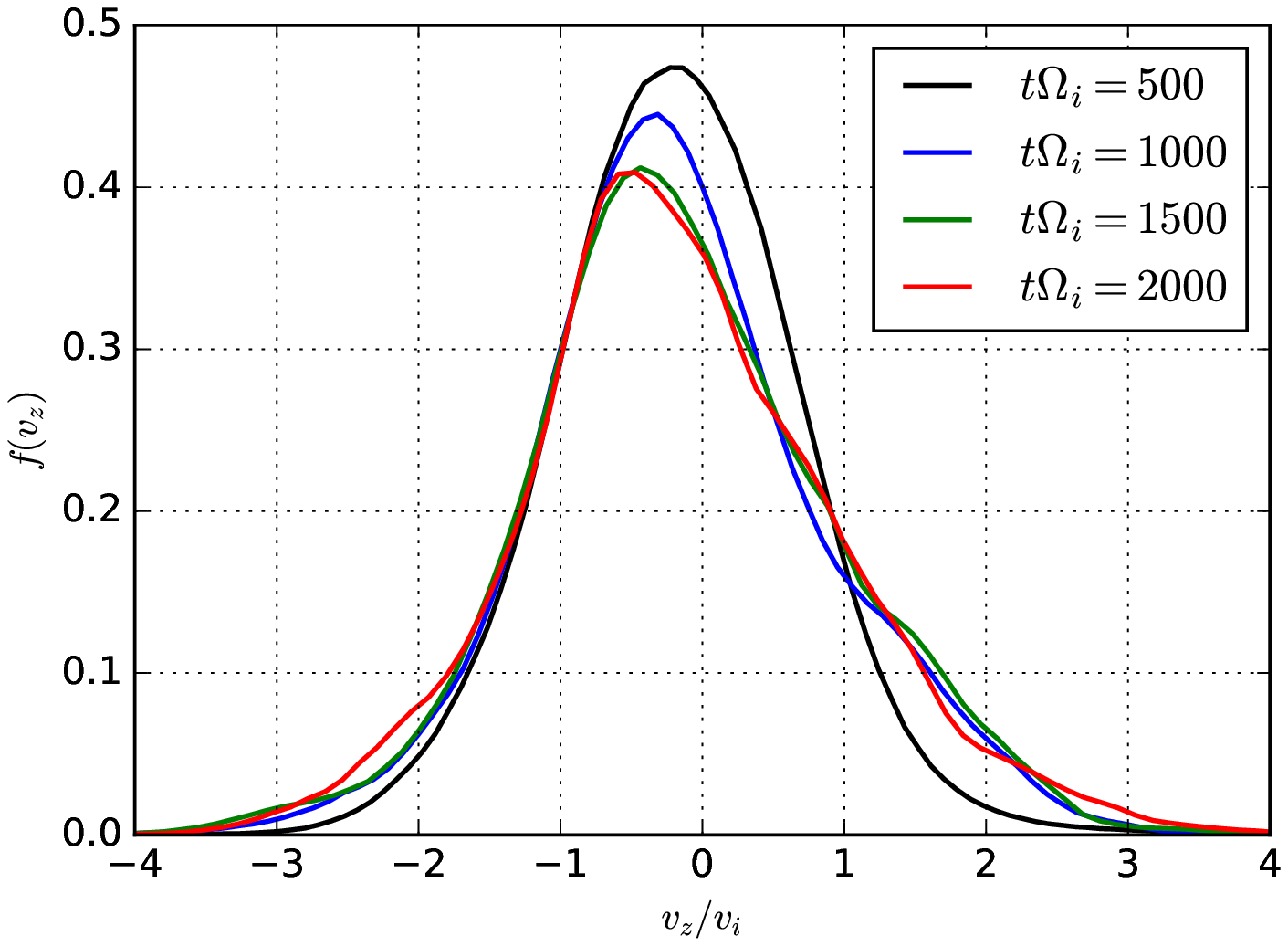}
   \caption{Ion velocity distribution $f(v_z)$ at various times in Run 1, where
     the velocity is normalized to the initial ion thermal velocity $v_i$ and
     $v_z\approx v_\parallel$.}
 \label{fig:dist}
 \end{figure}

Figure \ref{fig:temp} shows the surface contour of parallel ion temperature at
$t\Omega_i=1000$ in Run 1. Although the ion temperature distribution is also turbulent, temperature
enhancement structures are observed in the simulation with local peaks
$T_\parallel/T_{\parallel 0}\approx 10$ and parallel size ($\sim 25 d_i$) close to the wavelength of
the slow mode predicted by the PDI theory ($\sim 28 d_i$). 
Fourier power spectra of of the parallel ion temperature and density
fluctuation at $t\Omega_i=1000$ are shown in Figure \ref{fig:fft}a and Figure
\ref{fig:fft}b, respectively.
Clearly, the localized temperature enhancement structures in Figure
\ref{fig:temp} have a wave number
$k_\parallel L_z/2\pi = 17$, and they are associated with density fluctuation
near mode
(0,0,17) produced by PDI. Furthermore, there are strong oblique density fluctuations in long wavelength
($k_\parallel L_z/2\pi<5$, $k_\perp L_x/2\pi<5$ in Figures \ref{fig:fft}a and \ref{fig:fft}b) that are produced at the early
stage of the simulation when turbulence is developing. They are also present in
Run 2 (no injection), as shown in Figures \ref{fig:fft}c and \ref{fig:fft}d. Their heating effect causes the surface contour of
$T_\parallel$ appears localized in the perpendicular directions too.
The structures are
smoothed out later in the simulation, causing an overall heating of ions. 
Parallel velocity distributions ($v_\parallel\approx v_z$) of all ions in the simulation domain at
different stages are shown
in Figure \ref{fig:dist}. An initial Maxwellian velocity distribution
has been significantly changed after the injection of wave and subsequent PDI
through wave-particle interactions. Note that due to multiple Alfv\'en
waves loaded initially, ion velocity distribution does not center at $v_z=0$. Flattening of
the distribution function occurs near $v_z/v_i\approx1.0$ (where
$v_i=v_A\sqrt{\beta_i}$ is the initial thermal speed of ions) around
$t\Omega_i=1000$ and near $v_z/v_i\approx2.2$ around $t\Omega_i=2000$. A noticeable mount of
ions are accelerated to more than $2 v_i$. Since the ion acoustic
mode has a dispersion relation $\omega=k_\parallel
\sqrt{(3T_i+T_e)/m_{i}}=2k_\parallel v_i$ when $T_e=T_i$, the flattening of
ion velocity distribution is consistent with ion Landau damping of the slow
mode.

To study possible excitation of PDI in low-beta solar wind plasmas near 1 AU, we carry
out further simulations with conditions motivated by observations
\citep[e.g.][]{shi_apj_2015,huili_apj_2016}. Figure \ref{fig:history}b shows results of Run 5 and
6 which initially have $\beta_i=0.3$. When the background density perturbation drops to 
$\delta n/n \sim 0.1$ (note that $\delta n^2$ is multiplied by a factor of 2 in
the figure), we inject an Alfv\'en wave with amplitude $\delta B/B_0=0.3$. 
Similar to Run 1 with $\beta_i=0.01$, Run 5 shows features such as excitation of density fluctuations and parallel heating of ions after
injection, consistent with PDI. Although in a plasma with higher beta, a higher
amplitude Alfv\'en wave is needed to excite PDI (see Figure \ref{fig:linear}) and subsequent heating of ions is
weaker, PDI is still shown to be effective with parameters relevant to the solar wind near 1 AU.
The presence of slow waves will also likely change the turbulent energy cascade
in the parallel direction. All these effects suggest that turbulence and plasma
heating in low-beta plasma deserve considerable additional efforts.

\section{Discussion}
In this paper, we use 3D hybrid simulations to study the parametric decay
instability of a circularly polarized Alfv\'en wave in a turbulent low-beta plasma. 
It is shown that PDI is effective in such turbulence and the
pump wave 
decays into another Alfv\'en wave and an ion acoustic wave, which is evident in
the fluctuations of electromagnetic field and plasma density. The ion acoustic
wave is subsequently damped through Landau resonance, causing significant
heating of ions in the direction parallel to the background magnetic field.
In low-beta plasmas, PDI provides an
important channel for energy dissipation at a fluid scale, much larger than the
ion kinetic scales in traditional turbulence models.
It is often argued that parametric instabilities play a minor role in the
development of turbulence in magnetized collisionless plasmas. For example, 
\citet{howes_ptrsa_2015} argued that because of the anisotropic nature of plasma
turbulence, i.e. turbulent fluctuations satisfy $k_\parallel\ll k_\perp$, the
nonlinearity associated with Alfv\'en wave collisions (proportional to
$k_\perp$) dominate over parametric instabilities, whose growth rates are proportional to
$k_\parallel$. 
However, Eq. \ref{eq:pdi_growth} also shows that the growth rate of PDI
correlates inversely with $\beta$. In low-beta plasmas,
the growth of PDI can be faster than Alfv\'en wave collisions which are
independent of $\beta$.

PDI may also play an important role in plasma heating inside interplanetary
coronal mass ejections \citep{huili_apj_2016}, where plasma beta is observed to be on the order of 0.1
\citep{burla_ssr_1984}. Near the transition region of the solar
atmosphere, very low plasma beta ($10^{-4}-10^{-2}$) \citep{gary_sp_2001}
makes PDI a potentially robust mechanism for solar corona heating
\citep[e.g.,][]{prune_5soho_1997}.
We are expecting to see more direct evidence of PDI in the
solar wind close to the Sun, which will be made available by the upcoming NASA
Parker Solar Probe mission approaching as close as 8.5 solar radii from the Sun.

We have shown in a turbulent environment, PDI and Landau damping are still
effective, although their growth or damping rates are reduced compared to the prediction of
linear MHD theory or linear Vlasov theory in quiet plasmas. Turbulent fluctuations 
not only affect charged particle trajectories and background plasma density, invalidating the assumptions of
linear theories, but also introduce effective collisions between charged
particles in collisionless plasmas. Recently, \citet{versc_apj_2017} analyze
in-situ solar wind measurements and find that wave properties such as polarization of slow modes agree with MHD predictions better than the kinetic predictions,
suggesting that the plasma behaves more like a fluid. This interesting behavior
is worthy further studying using kinetic simulations, but it is beyond the scope
of this paper.

\begin{acknowledgments} 
The Los Alamos portion of
this research was performed under the auspices of the U.S. Department of
Energy. We are grateful for support from the LANL/LDRD program and DOE/OFES.
This research used resources provided by the Los Alamos National Laboratory
Institutional Computing Program, which is supported by the U.S. Department of
Energy National Nuclear Security Administration under Contract No.
DE-AC52-06NA25396. Contributions from VR were supported by NASA grants
NNX15AR16G and NNX14AI63G.
\end{acknowledgments}


 \end{document}